\documentclass[a4paper,Physsubmission, Phys]{SciPost}

\hypersetup{
    colorlinks,
    linkcolor={red!50!black},
    citecolor={blue!50!black},
    urlcolor={blue!80!black}
}

\usepackage[bitstream-charter]{mathdesign}
\DeclareSymbolFont{usualmathcal}{OMS}{cmsy}{m}{n}
\DeclareSymbolFontAlphabet{\mathcal}{usualmathcal}

\begin{document}

\begin{center}{\Large \textbf{
Recent ATLAS Results on Forward Physics and Diffraction\\
}}\end{center}

\begin{center}
Rafał Staszewski
\\on behalf of the ATLAS Collaboration
\end{center}

\begin{center}
Institute of Nuclear Physics Polish Academy of Sciences (IFJ PAN), Cracow, Poland\\

rafal.staszewski@ifj.edu.pl
\end{center}

\begin{center}
\today
\end{center}

\definecolor{palegray}{gray}{0.95}
\begin{center}
\colorbox{palegray}{
  \begin{tabular}{rr}
  \begin{minipage}{0.1\textwidth}
    \includegraphics[width=23mm]{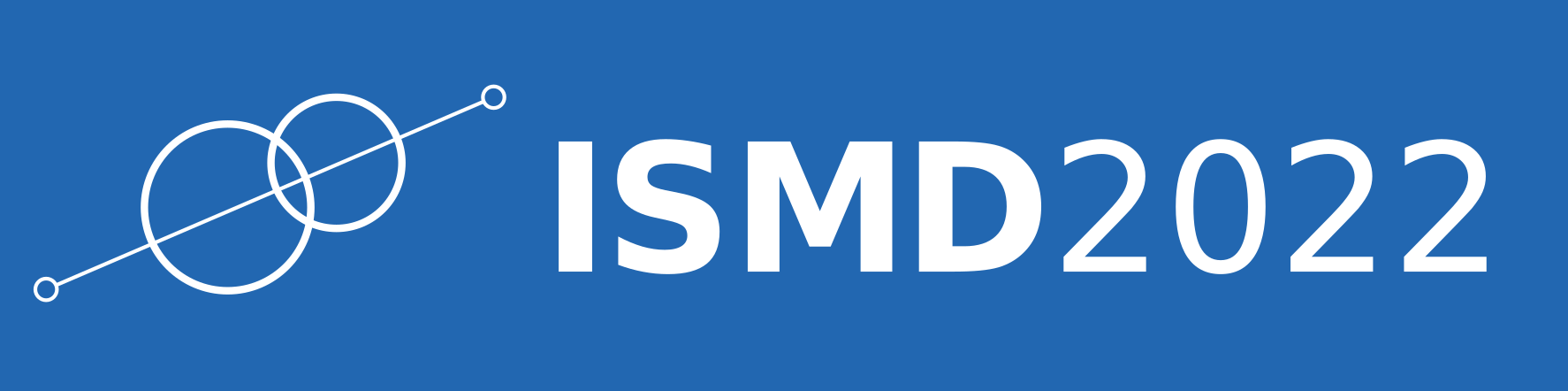}
  \end{minipage}
  &
  \begin{minipage}{0.8\textwidth}
    \begin{center}
    {\it 51st International Symposium on Multiparticle Dynamics (ISMD2022)}\\ 
    {\it Pitlochry, Scottish Highlands, 1-5 August 2022} \\
    \doi{10.21468/SciPostPhysProc.?}\\
    \end{center}
  \end{minipage}
\end{tabular}
}
\end{center}

\section*{Abstract}
{\bf
The ATLAS detector at the LHC is equipped with dedicated systems designed for the detection of forward protons produced in diffractive and photon-induced processes.
These detectors significantly extend the ATLAS physics reach.
Recent measurements performed using these forward proton detectors: elastic proton--proton scattering, exclusive charged pion-pair production, and two-photon production of lepton pairs are presented.
}
\let\thefootnote\relax\footnotetext{Copyright 2022 CERN for the benefit of the ATLAS Collaboration. Reproduction of this article or parts of it is allowed as specified in the CC-BY-4.0 license.}

\vspace{10pt}
\noindent\rule{\textwidth}{1pt}
\tableofcontents\thispagestyle{fancy}
\noindent\rule{\textwidth}{1pt}
\vspace{10pt}

\section{Introduction}
\label{sec:intro}
The ATLAS experiment \cite{atlas} at the LHC is used for studies of a wide range of physics topics. 
These include processes in which the interacting protons stay intact and are scattered at very small angles. 
ATLAS is equipped with two detector systems allowing measurements of such protons: ALFA \cite{alfa} (see also M. Schmidt's contribution in these proceedings) and AFP \cite{afp}.
Both systems are placed more than 200 m away from the interaction point and they use the Roman pot technology to detect scattered protons just a few millimetres from the LHC beam.

\section{Elastic scattering in the CNI region}

ALFA detectors were designed to study elastic scattering at the LHC energies.
The recent ATLAS measurement \cite{elas2500} was performed at 13~TeV centre-of-mass energy with special settings of the LHC magnetic fields characterised by a $\beta^\ast$ value of 2500~m.
Data taking in such conditions allows measurements of events with very small four-momentum transfer and gives access to the Coulomb-nuclear interference (CNI) region.

The analysis requires both scattered protons to be registered. 
The event selection is based on the correlations present in elastic events originating from the energy and momentum conservation: the positions of the two scattered protons measured in the detectors are anti-correlated, and the position of each proton is correlated to the elevation angle of its trajectory.
Such a selection results in a very clean event sample with background fraction below one permille.

The remaining background has two components.
The first one is due to a random coincidence of two protons registered in the same event.
Each of the two protons can be a part of the LHC beam halo, but could also be produced in a single diffractive $pp$ interaction ($pp\to pX$).
The shape of this background is estimated using data-driven templates based on events with a single proton registered.
The second source of background is the central diffractive process, where two protons remain intact, but a part of their energy is lost for the production of additional particles ($pp\to pXp$).
The shape of this background is modelled using Monte Carlo simulation.
The normalisation of both backgrounds is obtained in control regions using data.

In order to maximise the precision of the measurement, many of its ingredients were tuned using data-driven techniques.
These ingredients include:
\begin{itemize}
    \item detector alignment -- information necessary to correct for the limited and $t$-dependent detector acceptance,
    \item event reconstruction efficiency -- needed to correct for the detector inefficiency,
    \item modelling of the LHC beam optics -- used for the reconstruction of the scattering angle,
    \item detector resolution -- needed for the unfolding procedures,
    \item luminosity -- needed to convert the observed event distribution to the differential cross section.
\end{itemize}

\begin{figure}[htb]
\centering
\includegraphics[width=.6\textwidth]{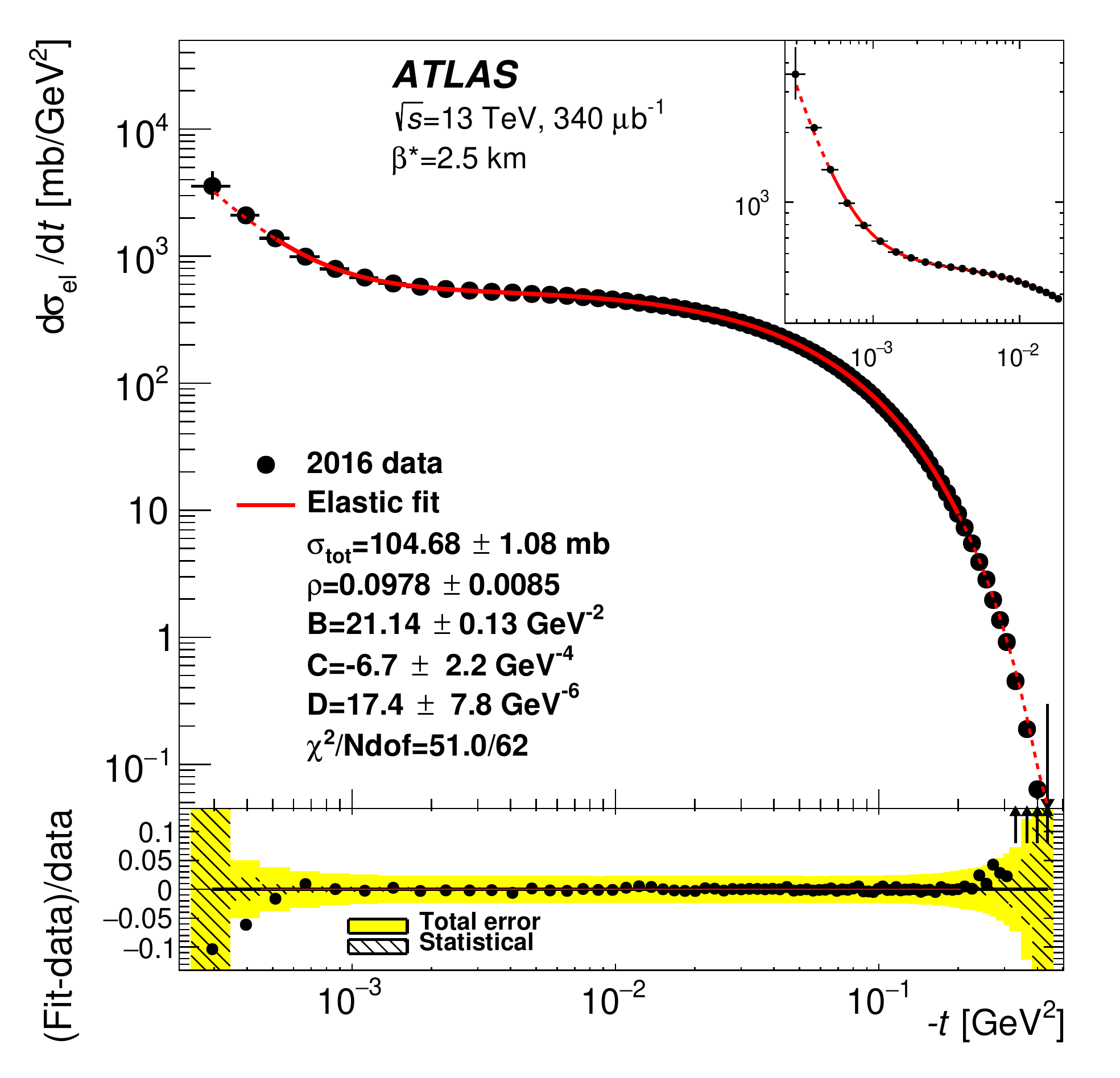}
\caption{Differential elastic cross section as a function of four-momentum transfer. From \cite{elas2500}.}
\label{fig:dsdt}
\end{figure}

The central result of the measurement is the differential elastic cross section as a function of the four-momentum transfer, $t$, corrected for experimental effects, with evaluated statistical and systematic uncertainties, see Figure \ref{fig:dsdt}.
The high-$|t|$ part of the distribution is dominated by the strong interactions.
The low-$|t|$ part is the CNI region, where the strong interactions are of similar magnitude as the electromagnetic interactions, and the measurement is sensitive to the interference effects.

\newcommand{\sigmatot}{\ensuremath{\sigma_\text{tot}}}

The measured differential elastic cross section can be used to extract several parameters of interest: $\sigmatot$ -- the total $pp$ cross section, $\rho$ -- the ratio of real to imaginary parts of the forward elastic scattering amplitude, $B$ -- the elastic slope, $C$ and $D$ -- higher order terms describing the slope.
This extraction is performed by fitting the data with the theoretical formula:
      \[
        \frac{\text{d}\sigma}{\text{d}t} = \frac{1}{16\pi}\left|f_{\text{N}}(t) + f_{\text{C}}(t)\text{e}^{\text{i}\alpha\phi(t)}\right|^2 ,
      \]
where
      \[
        f_{\text{C}}(t) = -8\pi\alpha\hbar c\frac{G^2(t)}{|t|},
        \qquad  
        f_{\text{N}}(t)  =  \left(\rho + \text{i}\right)\frac{\sigmatot}{\hbar c}\text{e}^{(-B|t| - C|t|^{2} - D|t|^{3})/2},
      \]
$\phi(t)$ is the Coulomb phase, and $G(t)$ is the proton form factor.
The fitting procedure used in the analysis took into account both the statistical and systematic uncertainties together with the correlations between their values in different measurement intervals.

\begin{figure}[htb]
\includegraphics[width=0.5\textwidth]{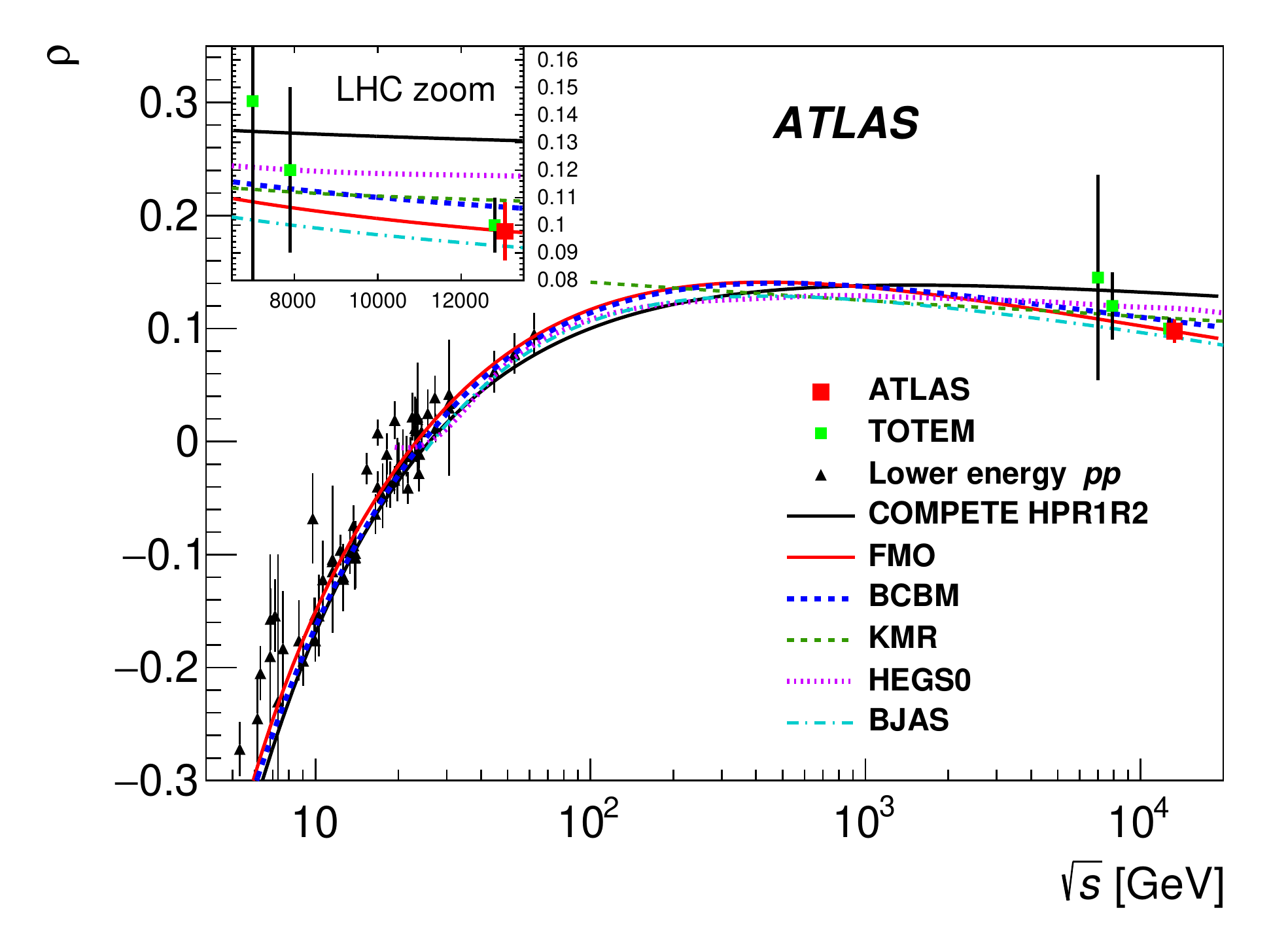}
\includegraphics[width=0.5\textwidth]{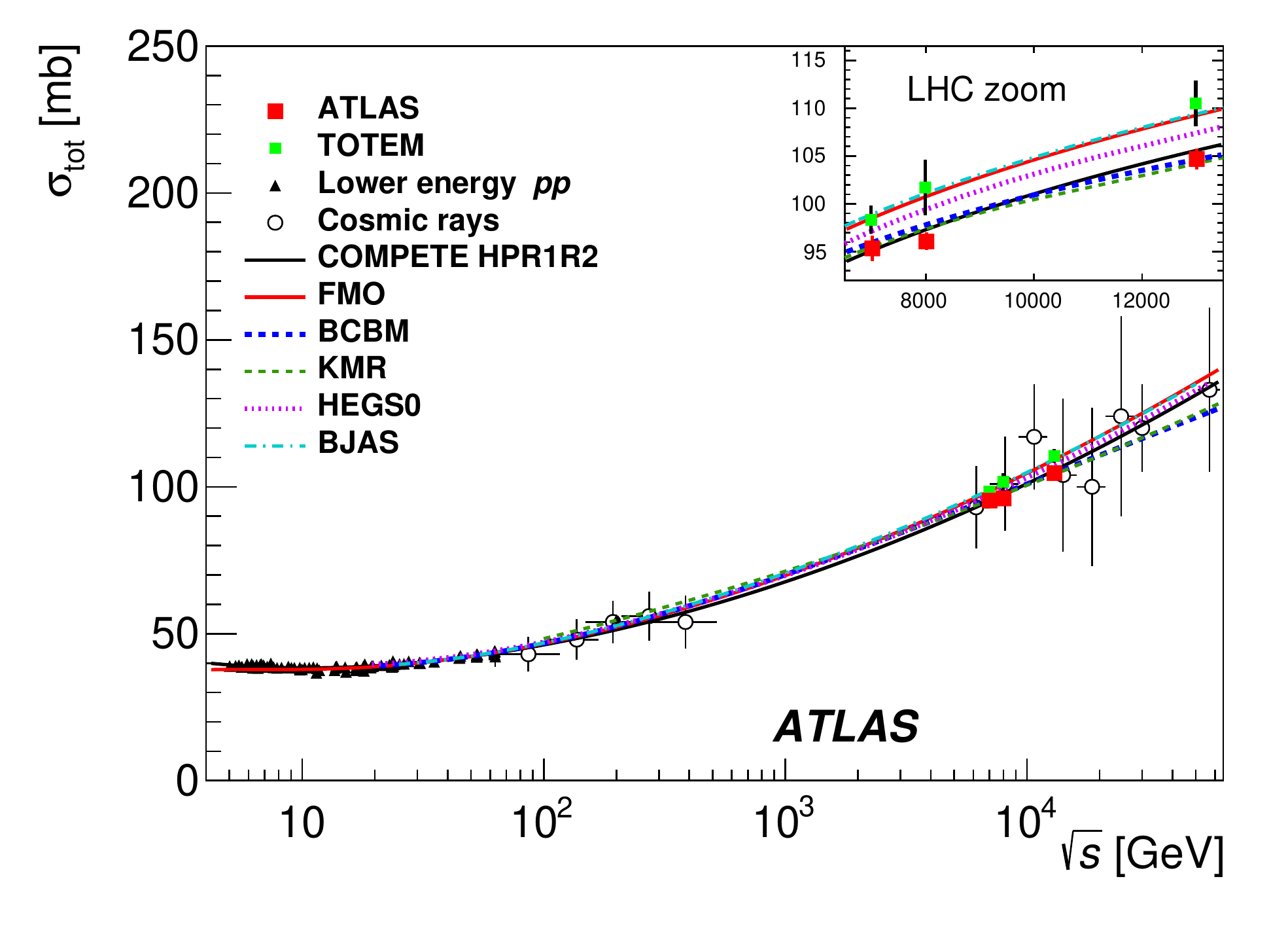}
\caption{Energy evolution of the $\rho$ parameter (left) and total cross section (right). From \cite{elas2500}.}
\label{fig:rho_stot}
\end{figure}

Figure \ref{fig:rho_stot} (left) presents the result of the present $\rho$-parameter measurement together with other available data as a function of the centre-of-mass energy accompanied by the predictions of several theoretical models.
An important observation is that the measured value disagrees with the pre-LHC COMPETE calculations \cite{compete}.
This theoretical model assumed no asymptotic difference between $pp$ and $p\bar p$ interactions and that the present rate of growth of \sigmatot{} with energy will continue. 
The ATLAS results can therefore be interpreted as a disproval of one of these assumptions. 

Another important result is the value of the total $pp$ cross section, see Figure \ref{fig:rho_stot} (right).
This is the most precise measurement of \sigmatot{} at this energy, which was possible in particular thanks to the dedicated ATLAS luminosity measurement performed for these data.
One can observe that the discrepancy between the ATLAS and TOTEM measurements, present already at 7 TeV and 8 TeV, still persists (the ATLAS  value is about 5\% below the one reported by TOTEM).
It is interesting to notice that some theoretical descriptions prefer the ALFA results, while others agree better with the TOTEM values.

\section{Exclusive production of charged pion pairs}

While the ALFA detectors were designed with the primary purpose of measuring elastically scattered protons,
they can be used for studies of processes with more complex final states.
Recently, ATLAS has conducted a measurement of the exclusive production of charged-pion pairs, $pp \to p \pi^+ \pi^- p$.
The analysed data were collected at the centre-of-mass energy of 7~TeV and with $\beta^\ast$ value of 90 m.
The measurement required both protons to be reconstructed using the ALFA detectors, which makes it the first fully exclusive measurement of this process at the LHC.
It was possible because the data from the ALFA detectors were collected together with the data from all other ATLAS subsystems, in particular the Inner Detector (ID), used to measure the pion tracks.

\begin{figure}[ht]
\includegraphics[width=0.5\textwidth]{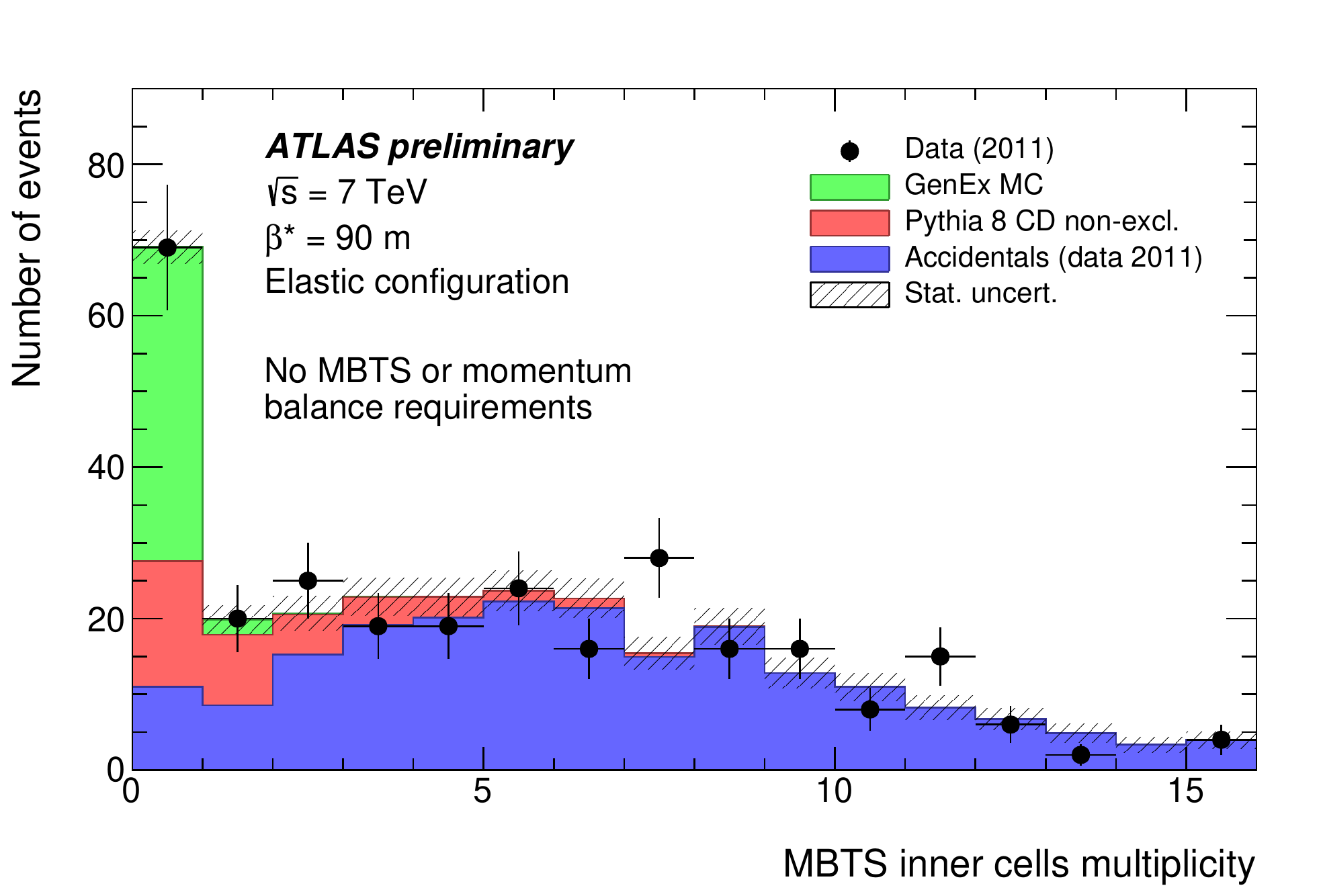}
\includegraphics[width=0.5\textwidth]{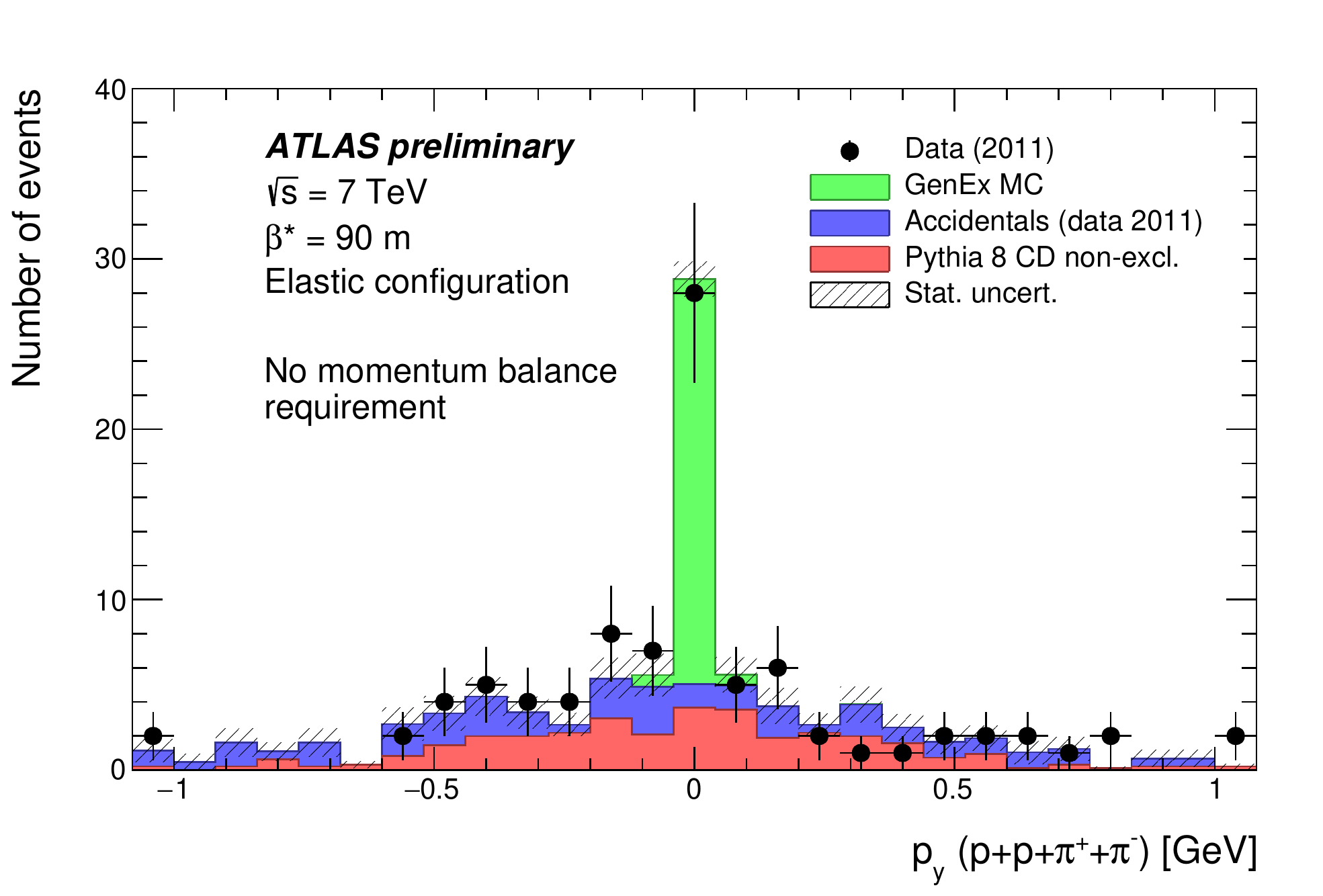}
\caption{Left: distribution of the MBTS signal. Right: distribution of the total vertical momentum of the two protons and two pions.}
\label{fig:pipi}
\end{figure}

In events where the final state consists of only four particles -- two protons measured in ALFA and two pions in the ID -- no additional particles should be present.
This exclusivity of the measured events was ensured in two ways.
First, the signal from the MBTS detector \cite{mbts} was used, see Figure \ref{fig:pipi} (left).
For signal events, no particles should be passing through the MBTS and its response should be close to zero.
For background events, the MBTS may register particles that were not observed in the tracking detectors.

The second exclusivity condition is based on the transverse momentum balance.
For signal events, the total transverse momentum of the $p\pi\pi p$ system should be close to zero.
For events where additional particles were produced but not measured, the $p_\text{T}$ imbalance may be larger, see Figure \ref{fig:pipi} (right).

The measurement was performed for two event topologies defined based on the signs of the vertical momenta of the two protons:  elastic-like -- with $p_{y,1} \cdot p_{y,2} < 0$, and  anti-elastic -- with $p_{y,1} \cdot p_{y,2} > 0$.
The measured fiducial cross sections are:
  \[ \sigma_\text{elastic} = 4.8 \pm 1.0 \text{(stat)} ^{+0.3}_{-0.2} \text{(syst)} \pm 0.1 \text{(lumi)} \pm 0.1 \text{(model)}\ \text{µb},\]
  \[ \sigma_\text{anti-elastic} = 9 \pm 6 \text{(stat)} \pm 1 \text{(syst)} \pm 1 \text{(lumi)} \pm 1 \text{(model)}\ \text{µb}.\]

\section{Two-photon production of lepton pairs}

One drawback of the ALFA detectors is that for a good performance, they require high-$\beta^\ast$ LHC optics.
This limits their physics reach to processes with relatively high cross section.
On the other hand, the AFP detectors were designed to operate in standard LHC conditions, with low $\beta^\ast$ values (below 1 m) and high luminosity.
This allows measurements of processes with much smaller cross sections, such as two-photon exclusive production.
The first result obtained with the AFP detectors concerned the exclusive two-photon production of lepton pairs: $pp\to pl\bar l p$ \cite{afp2l}.

\begin{figure}[ht]
\centering
\includegraphics[width=0.7\textwidth]{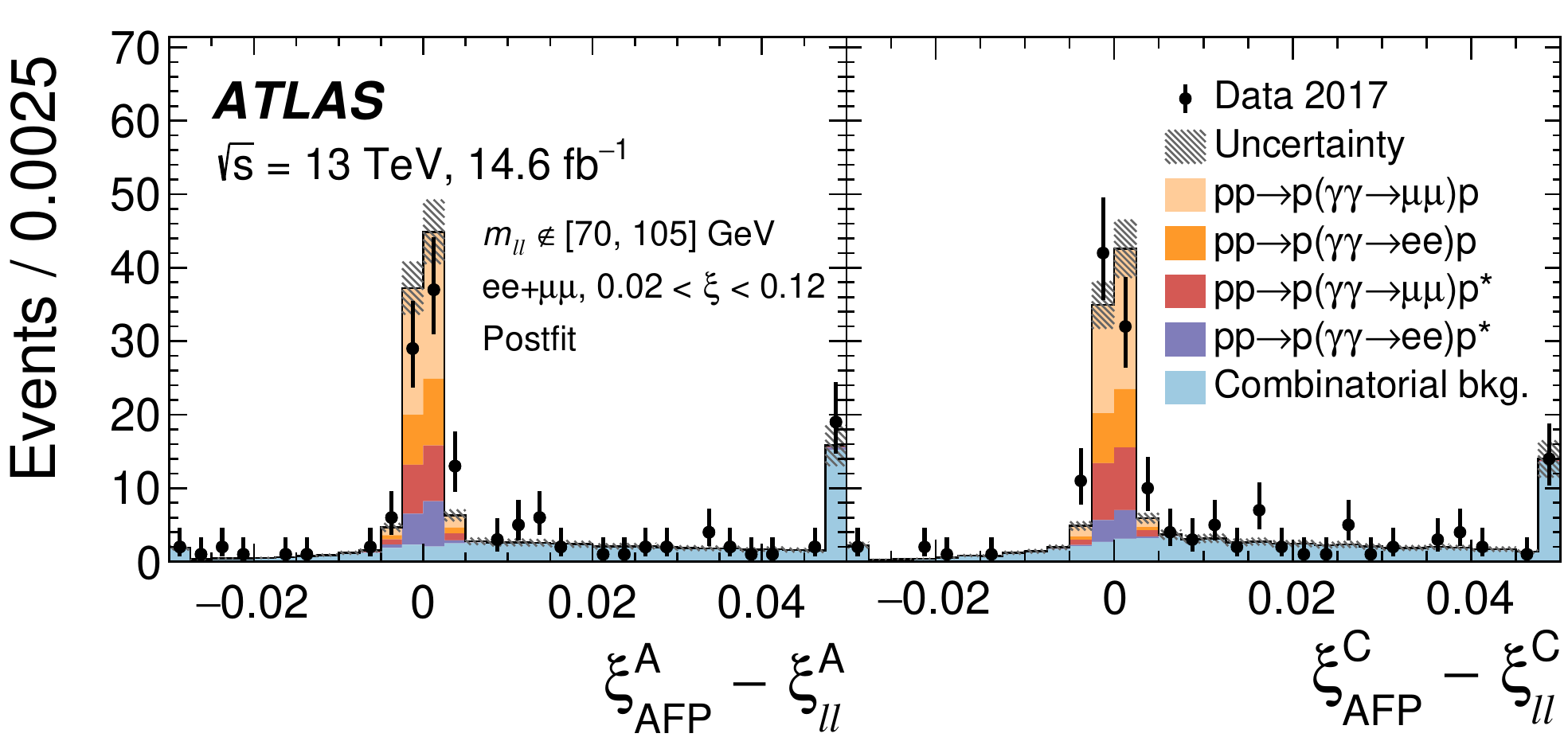}
\caption{Distribution of the difference of the relative momentum loss of a forward proton measured directly and calculated from the momenta of the measured leptons. The two panels show distributions for events with the forward proton detected on two sides of the interaction point. From \cite{afp2l}.}
\label{fig:dxi}
\end{figure}

The analysis used several methods for rejecting non-exclusive background. 
The requirements of no additional tracks pointing to the interaction vertex and small dilepton acoplananarity were similar to previous measurements without proton tagging, for example \cite{notag}.
The measurement of the forward proton allowed using an additional exclusivity requirement.
It originates from the fact that in the signal events the proton momentum can be calculated from the measurement of the central dilepton system. 
The difference between the calculated value and a direct AFP-based measurement is expected to be small for signal events.
The background originates from a random coincidence of two $pp$ interactions: one leading to the production of a lepton pair (for example the Drell-Yan process) and the other resulting in a forward proton (mainly single diffraction). 
Such a background can be easily modelled by a random mixing of data events with leptons and with protons.
Because the two above methods applied for background events measure properties of two independent $pp$ interactions, the distribution of their difference will be much flatter than for the signal events; see Figure \ref{fig:dxi}.

The main results of the measurement are the fiducial cross sections in the electron and muon channel:
  \[\sigma_{ee} = 11.0 \pm 2.6 \text{(stat.)}  \pm 1.2 \text{(syst.)} \pm 0.3 \text{(lumi.) fb}\]
  \[\sigma_{\mu\mu} = 7.2  \pm 1.6 \text{(stat.)} \pm 0.9 \text{(syst.)} \pm 0.2 \text{(lumi.) fb}\]
These results can be compared to theoretical calculations in order to validate the present understanding of the rescattering corrections that affect the cross sections for two-photon processes.

\section{Conclusions}

ATLAS detectors measuring forward protons, ALFA and AFP, proved to be an invaluable part of the experiment, significantly enhancing its physics program.
This paper presented three flagship measurements: elastic scattering, exclusive pion-pair production, and exclusive photon-induced lepton-pair production.
The results are sensitive to the dynamics of long-range strong interactions, where perturbative QCD methods cannot be applied, and await further phenomenological investigations.
Other interesting measurements are possible with the data already collected during LHC Run~2.
In addition, both systems plan their operation also in Run~3.

\paragraph{Funding information}
This work is supported in part by the Polish Ministry of Education and Science project no. 2022/WK/08.

\nolinenumbers

\end{document}